\documentclass[aps,prl,twocolumn,groupedaddress]{revtex4}
\usepackage{graphicx}
\newcommand{\G}{$G(0)$}
\newcommand{\Gab}{$^{71}$Ga}
\newcommand{\Gaa}{$^{69}$Ga}
\newcommand{\As}{$^{75}$As}

\begin{document}

\title{Spin Transition in Strongly Correlated Bilayer Two Dimensional Electron Systems}

\author{I.~B. Spielman$^1$, L. A. Tracy$^1$, J.~P. Eisenstein$^1$, L.~N. Pfeiffer$^2$, and K.~W. West$^2$}

\affiliation{$^1$California Institute of Technology, Pasadena CA 91125 
\\
	 $^2$Bell Laboratories, Lucent Technologies, Murray Hill, NJ 
07974\\}

\date{\today}

\begin{abstract}
Using a combination of heat pulse and nuclear magnetic resonance techniques we demonstrate that the phase boundary separating the interlayer phase coherent quantum Hall effect at $\nu_T = 1$ in bilayer electron gases from the weakly coupled compressible phase depends upon the spin polarization of the nuclei in the host semiconductor crystal. Our results strongly suggest that, contrary to the usual assumption, the transition is attended by a change in the electronic spin polarization.  
\end{abstract}


\maketitle

A remarkable quantum fluid emerges from bilayer two-dimensional electron systems (2DES) in perpendicular magnetic fields $B$ when the layer separation is small and the total density of electrons $N_T$ in the bilayer equals the degeneracy $eB/h$ of a single spin-resolved Landau level produced by the field.  Inter- and intralayer Coulomb interactions are of comparable strength in this fluid and result in spontaneous interlayer quantum coherence among the electrons in the system.  The system may be viewed in several equivalent ways, including as a pseudospin ferromagnet\cite{perspectives} or a superfluid of interlayer excitons\cite{eisenmac}.  In addition to exhibiting the integer quantized Hall effect (QHE) when parallel currents flow in the two layers, this collective state displays a number of other very unusual transport properties, including Josephson-like interlayer tunneling\cite{spielman1} and a diverging conductivity for counterflowing currents in the two layers as the temperature is reduced toward zero\cite{kellogg1,tutuc1}.  

As the layer separation is increased, the excitonic phase first weakens and then gives way to a non-QHE, weakly-coupled phase lacking interlayer coherence\cite{murphy}. When the layer separation is very large this phase consists of two independent 2D electron systems. For equal layer densities, each 2DES is at Landau level filling fraction $\nu = 1/2$ and is well-described as a metallic state of composite fermions\cite{bert}.  Closer to the critical layer separation the situation is much less clear.  Recent experiments have revealed a strong enhancement of interlayer drag\cite{kellogg2} in the vicinity of the transition and that the critical layer separation increases when small anti-symmetric layer density imbalances are imposed\cite{tutuc2,spielman2}.  Although these findings are consistent with recent theoretical work\cite{stern,hanna,joglekar}, the precise nature of the transition is unknown.  Fundamental questions, such as the order of the transition, how many phases actually exist, and what their electronic structures are near the critical point(s), remain unanswered. 

A common simplifying assumption has been that the electron spins in the bilayer 2DES at $\nu_T = 1$ are frozen out by the Zeeman energy. While this is perhaps reasonable in the gapped excitonic phase at small layer separation, at large separation it conflicts with the several reports of incomplete polarization at $\nu = 1/2$ in single layer 2D systems at low density\cite{igor,barrett,levy}.  Given the poor current understanding of the phase transition between the excitonic superfluid and the non-QHE phases at $\nu_T = 1$, this question of spin configuration looms large.  Here we report compelling evidence that the spin degree of freedom is active in this transition and conclude that near the critical point the spin polarization of the excitonic state must exceed that of the competing non-QHE phase. 

The spin of an electron in the bilayer 2DES is coupled to the nuclear spin polarization in the $\rm GaAs/Al_xGa_{1-x}As$ heterostructure sample via the hyperfine contact interaction. The effective electronic Zeeman energy is $E_Z = -g\mu_B(B-B_N)$, where $g \approx -0.44$ is the $g$-factor of electrons in GaAs and $B_N$ is the effective magnetic field produced by the nuclear polarization.  The nuclear field $B_N$, which can reach $\sim 5.3$ T if the nuclear polarization is complete, appears with a minus sign due to the negative $g$-factor of electrons near the $\Gamma$-point of the GaAs conduction band. In the present experiments the nuclear polarization, and hence the electronic Zeeman energy, is controlled via heat pulse and NMR techniques. 

The bilayer sample used here consists of two 18 nm GaAs quantum wells separated by a 10 nm $\rm Al_{0.9}Ga_{0.1}As$ barrier layer.  In its as-grown state each quantum well contains a 2DES with density $4.4\times10^{10}$ $\rm cm^{-2}$ and a low temperature mobility of about $9.3\times10^5$ $\rm cm^2/Vs$.  A square mesa, 250 $\mu$m on a side, with four arms extending outward to ohmic contacts is patterned onto the sample. Metallic gates covering the front and thinned backside of the sample's central region allow independent control of electron densities in each quantum well.  Conventional resistance and interlayer tunneling conductance measurements are performed using techniques described in detail elsewhere\cite{spielman1,kellogg1}.  The sample is suspended in vacuum by Au wires. These wires provide both electrical and excellent thermal contact between the sample and the cold finger of a dilution refrigerator; thermal relaxation times of just a few seconds are observed at 50 mK. The sample is surrounded by a rectangular 8-turn NMR coil for applying radio-frequency magnetic fields parallel to the plane of the 2DES. A small resistive heater is attached directly to the 50 $\mu$m-thick sample for heat pulse experiments. 

As reported previously\cite{spielman1}, the coherent excitonic phase at $\nu_T = 1$ is readily distinguished from the compressible phase at larger effective layer separation by the interlayer tunneling conductance at zero bias, $G(0)$.  In the compressible phase \G\ is heavily suppressed by Coulomb blockade-like effects occuring within the individual layers\cite{eisenstein1}.  In contrast, in the coherent excitonic phase a very sharp peak in the tunneling conductance appears at zero bias.  The magnitude of this peak grows continuously as the effective layer separation is reduced, eventually dwarfing all other features in the tunnel spectrum.  This feature, which reflects the existence of a Goldstone collective mode in the coherent state, allows for accurately locating the phase boundary. In the present sample, under equilibrium conditions, the phase boundary occurs at an effective layer separation of $d/\ell = 1.97$, where $d$ = 28 nm is the nominal quantum well center-to-center separation and $\ell = (\hbar/eB)^{1/2}$ is the magnetic length at $\nu_T = 1$\cite{layers}. 

\begin{figure}
\centering
\includegraphics[width=3.2 in, bb=176 148 427 428]{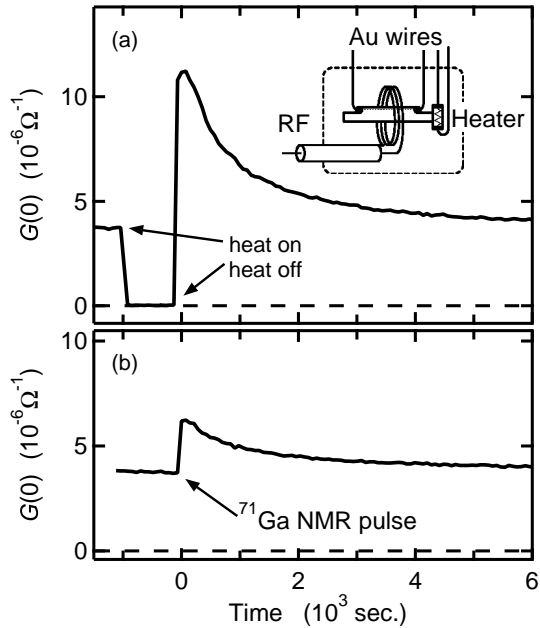}
\caption{\label{fig:fig1}Enhancement of zero bias tunneling conductance \G\ at $\nu_T = 1$ in response to heat and \Gab\ NMR pulses at $d/\ell = 1.92$ and $T = 35$ mK.  Inset depicts experimental set-up.}
\end{figure}

Figure 1a shows the response of \G\ at $\nu_T = 1$ to a 900 sec. heat ``pulse'' applied to the sample.  Prior to the pulse the system is just inside the coherent phase at $d/\ell = 1.92$ and $T = 35$ mK.  The conductance at zero bias is significant: \G\ = $3.7 \times 10^{-6} ~\Omega^{-1}$.  Sufficient power is then applied to the on-chip heater to raise the sample temperature (but not the dilution refrigerator's) to about 350 mK.  This destroys the zero bias tunneling peak and \G\ falls to zero.  When the heat is removed (at $t$ = 0 in the figure) the electron system rapidly cools.  Remarkably, as this cooling proceeds (in well less than a minute) the tunneling conductance does not return to its initial value, but to one several times larger, suggesting that the coherent phase has been somehow strengthened by the heating and cooling process. Only after several thousand seconds does \G\ return close to its pre-heat pulse equilbrium value.

One candidate for the long time constant for relaxation of the tunneling conductance following the heat pulse is the nuclear spin-lattice relaxation time, $T_1$.  To investigate the possible involvement of nuclear spins, radio-frequency (RF) magnetic fields were applied to the sample using the coil described above.  Swept-frequency measurements clearly demonstrate a sharply resonant response\cite{reswidth} of the tunneling conductance at the appropriate NMR frequencies for all three relevant nuclei in the sample: \As, \Gaa, and \Gab.  Figure 1b shows the response, again at $\nu_T = 1$, $d/\ell = 1.92$, and $T = 35$ mK, of \G\ to a short RF pulse\cite{RFpulse} at 40.175 MHz. This is appropriate for NMR of \Gab\ at $B = 3.11$ T. The tunneling conductance immediately increases in response to the RF and then decays slowly after the RF is removed, the decay time being the same as that seen in the heat pulse experiment. Very similar responses are observed for NMR of \Gaa\ and \As.  Note that unlike a heat pulse which warms the entire sample, NMR pulses heat only the nuclear spins and thus do not initially quench the tunneling conductance.

The data in Fig. 1 make clear that both heat and RF pulses temporarily enhance the zero bias tunneling conductance in the excitonic phase. In fact, as Fig. 2 demonstrates, a complete zero bias tunneling peak can be created by RF or heat pulses even when no such peak is present in equilibrium.  Figs. 2a and 2b show the tunneling conductance $dI/dV$ vs. interlayer voltage $V$ at $\nu_T = 1$, $T = 35$ mK, and $d/\ell = 1.976$, before and shortly after a heat pulse.  Before the heat pulse the tunneling conductance shows the deep minimum at zero bias representative of the compressible phase.  After the pulse, but before complete equilibrium is re-established, a zero bias peak is clearly seen, revealing the presence of excitonic fluid.  The peak gradually disappears, with time-constant comparable to those seen in Fig. 1.  

\begin{figure}
\centering
\includegraphics[width=3.2 in, bb=135 263 420 581]{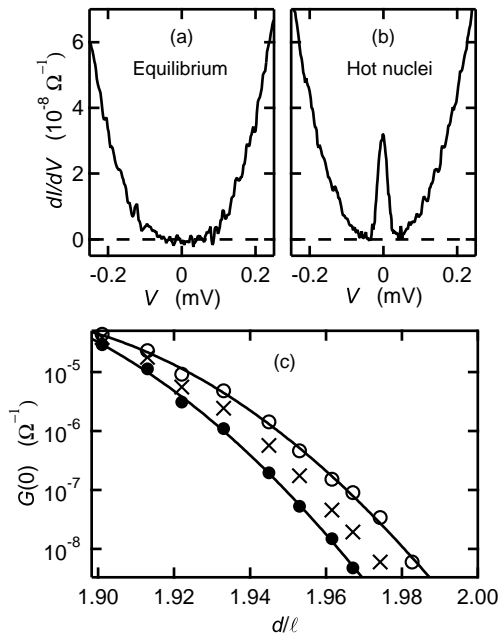}
\caption{\label{fig:fig2}(a) Equilibrium tunneling conductance vs. interlayer voltage around zero bias at $\nu_T =1$ and $d/\ell = 1.976$ and $T = 35$mK. No zero bias peak is present. (b) Same as (a), except data taken immediately after a heat pulse. Zero bias peak now present. (c) Summary of zero bias conductance at $\nu_T = 1$ vs. $d/\ell$ in equilibrium (solid dots), after \Gab\ NMR pulses (crosses), or after heat pulses (open dots).}
\end{figure}

Fig. 2c summarizes the effect of heat and \Gab\ NMR pulses on \G\ at $T = 35$ mK.  The solid dots show the $d/\ell$ dependence of \G\ measured under equilibrium conditions.  The crosses and open dots do the same, but from measurements taken shortly after NMR and heat pulses respectively.  At all $d/\ell$, heat and NMR pulses enhance \G, with the effect becoming proportionally larger close to the critical point.  The heat pulses always produce a larger effect than the NMR pulses.

A simple model which consistently explains the data in Figs. 1 and 2 invokes a competition between two electronic phases with differing electronic spin polarization.  The NMR results prove that nuclear spins are involved in this competition.  This is sensible since the hyperfine interaction couples the nuclear spin polarization to the electronic spin Zeeman energy. At $T = 35$ mK and $B = 3$ T the equilibrium nuclear polarization in GaAs is about 2.5, 3.5,  and 4.5\% for \As, \Gaa, and \Gab, respectively.  Together these polarizations produce an effective nuclear magnetic field $B_N \approx 0.17$ T.  This field $\it reduces$ the electronic Zeeman energy slightly.  Both heat and resonant RF pulses reduce the nuclear polarization and thereby reduce $B_N$.  This temporarily $\it increases$ the Zeeman energy and thus favors the electronic phase with larger electronic spin polarization. Our results demonstrate that this is the excitonic QHE phase. Heating is more effective than NMR simply because all three nuclear species are depolarized simultaneously.  

Recent experiments\cite{kellogg2} have been cited as evidence that in the vicinity of the phase transition static density fluctuations in the sample lead to phase separation at $\nu_T = 1$, with regions of coherent excitonic phase co-existing with regions of weakly-coupled background fluid\cite{stern}.  As $d/\ell$ is reduced the fraction $f$ of the sample containing the excitonic fluid increases. There is thus no single critical effective layer separation, but instead a range of values dependent upon the level of density, and possibly other structural fluctuations in the sample.  The present findings demonstrate that $f$ depends upon the electronic Zeeman energy difference between the two phases.  By destroying the nuclear polarization, heat and NMR pulses increase the Zeeman energy and thereby increase $f$, even rendering it finite when it is zero in equilibrium.  Since it is natural to assume that the zero bias tunneling conductance is proportional to $f$, this model offers a ready explanation for the increase in \G\ following heat and/or NMR pulses.  The effect is temporary, lasting only until the nuclear spins return to thermal equilibrium.  

\begin{figure}
\centering
\includegraphics[width=3.2 in, bb=125 127 408 444]{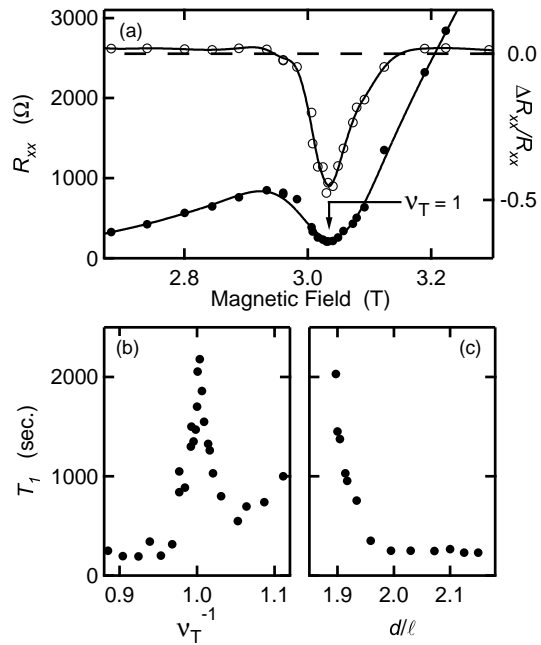}
\caption{\label{fig:fig3} (a) Resistively-detected NMR.  Solid dots: Equilibrium longitudinal resistance $R_{xx}$ near $\nu_T =1$. Open dots: Fractional change in $R_{xx}$ in response to \Gab\ NMR pulses. (b) NMR $T_1$ relaxation time vs. inverse filling factor ${\nu_T}^{-1}$. (c) $T_1$ vs. $d/\ell$ at $\nu_T = 1$.  All data at $T = 35$ mK.}
\end{figure}

The nuclear spin-lattice relaxation time $T_1$ should depend strongly on the electronic density of states within the GaAs quantum wells and thus should be sensitive to whether the gapped excitonic phase or the gapless non-QHE phase is present.  To demonstrate this, we now turn to measurements of the longitudinal resistance $R_{xx}$ of the bilayer 2DES.  We have found that $R_{xx}$ at $\nu_T = 1$ displays a response to heat and RF pulses completely consistent with the tunneling results described previously. Figure 3a shows $R_{xx}$ and the fractional change $\Delta R_{xx}/R_{xx}$ in response to RF pulses tuned to the \Gab\ NMR line, as a function of magnetic field at $T = 35$ mK.  For these data $\nu_T = 1$ occurs at $B =3.03$ T where the effective layer separation is $d/\ell = 1.90$.  The deep minimum in $R_{xx}$ seen at this field demonstrates the presence of the excitonic QHE phase.  Depolarizing the \Gab\ nuclei with an RF pulse temporarily deepens this minimum substantially: At $\nu_T = 1$ $R_{xx}$ falls by nearly a factor of 2. This strong reduction of $R_{xx}$ is confined to filling factors within a few percent of $\nu_T = 1$. Within the phase separation model described above, the NMR pulse-induced drop in $R_{xx}$ results from the temporary increase in $f$, the fraction of the sample occupied by the excitonic phase. Increasing $f$ improves the electrical connectivity of the QHE phase across the sample and thereby reduces the longitudinal resistance.

Unlike the detection of NMR via the tunneling conductance, we find that $R_{xx}$ remains sensitive, albeit weakly, to NMR and heat pulses well away from the excitonic phase.  This allows us to examine the behavior of $T_1$ in both the excitonic and weakly-coupled phases. Figures 3b and 3c reveal that the relaxation time\cite{T1} at $T = 35$ mK is, as expected, very sensitive to the presence of the excitonic phase at $\nu_T = 1$.  Figure 3a shows that $T_1$ exceeds 2000 sec. at $\nu_T = 1$ and $d/\ell = 1.90$, but rapidly falls on moving away from this filling factor.  Figure 3c shows that $T_1$ at precisely $\nu_T = 1$ also falls rapidly when the phase boundary between the excitonic and compressible phases is crossed by increasing $d/\ell$.

Our findings imply that the electronic spin polarization of the weakly-coupled phase at $\nu_T = 1$ cannot be complete.  This agrees with previous reports\cite{igor,barrett,levy} that the polarization of a single-layer 2DES at $\nu = 1/2$ is incomplete when the electron density is sufficiently low.  Within the composite fermion model, a partially polarized $\nu = 1/2$ state possesses two Fermi surfaces, one for each spin species\cite{bert}. Low energy electron spin-flip scattering processes are therefore possible and these should lead to relatively rapid, Korringa-like, nuclear spin-lattice relaxation.  This is consistent with the relative short $T_1$ times for $d/\ell > 2$ that are shown in Fig. 3c.  Although not shown, at these large effective layer separations we find that $T_1$ rises as the temperature $T$ is reduced, although less rapidly than the usual Korringa law, $T_1T =$ constant, would suggest. Unlike the situation close to the phase boundary, the origin of the NMR pulse-induced $\Delta R_{xx}$ at large $d/\ell$ is not well understood. 

The $T_1$ time in the excitonic phase is quite slow.  Whether those nuclear spins which are surrounded by the excitonic QHE phase relax via some, as yet unknown, low-lying spin-flip excitations within the same phase or by spin diffusion towards regions of the sample containing the weakly-coupled, short $T_1$, fluid is an interesting question for future work.  Equally interesting are the questions of whether an NMR or heat pulse-induced response should be detected at small $d/\ell$, deep in the excitonic phase where $f$ is close to unity, and whether the excitonic phase is in fact $fully$ polarized at $T = 0$.

In summary, we have demonstrated that the competition between the excitonic QHE and weakly-coupled compressible phase of bilayer 2D electron systems at $\nu_T = 1$ depends upon the degree of nuclear spin polarization in the sample.  This observation strongly suggests that the spin polarization of the bilayer 2D electron system increases when the transition from weakly-coupled to excitonic QHE phase at $\nu_T = 1$ occurs. 

We thank S.E. Barrett, S. Das Sarma, E. Demler, S.M. Girvin, B. Halperin, A.H. MacDonald, and A. Stern for helpful discussions. This work was supported by the NSF under Grant No. DMR-0242946 and the DOE under Grant No. DE-FG03-99ER45766.


\begin{thebibliography}{99}
\bibitem{layers} This critical $d/\ell$ value is about 7\% larger than that seen in other samples with nominally identical structures.  We attribute this discrepancy to imprecise knowledge of the layer thicknesses in the sample.  The apparently larger $d/\ell$ value suggests that the layers are thinner than expected.  This is consistent with the higher levels of tunneling observed in this sample.
\bibitem{reswidth} We find NMR linewidths of approximately 20 kHz.
\bibitem{T1} Here $T_1$ is defined as the time required for the heat or NMR-induced increase in resistance, $\Delta R_{xx}$ to fall to $e^{-1}$ of its initial value.
\end{thebibliography}
\end{document}